\documentclass[11pt,twoside]{article}

\usepackage{asp2006}
\usepackage{epsf}
\usepackage{psfig}
\usepackage{lscape}

\begin{document}
\title{An approach to the Riemann problem for SPH inviscid ideal flows: consequences for the state equation}   
\author{G. Lanzafame}   
\affil{INAF - Osservatorio Astrofisico di Catania, Via S. Sofia 78 - 95123 Catania - Italy}    

\begin{abstract} 
In the non viscous fluid dynamics, Smooth Particle Hydrodynamics (SPH), as a free Lagrangian "shock capturing" method adopts either an artificial viscosity contribution or an appropriate Riemann solver technique. An explicit or an implicit dissipation, introduced in such techniques, is necessary to solve the Euler equations to solve flow discontinuities (the Riemann problem). Dissipation is useful to smooth out spurious heating and to treat transport phenomena. A simple, effective solution of the Riemann problem is here proposed, based on an empirical reformulation of the equation of state (EoS) in the Euler equations in fluid dynamics, whose limit for a motionless gas coincides with the classic EoS of ideal gases. Results on 1D shock tube tests are here shown, as well as a 3D transport application on accretion discs in close binaries (CBs).
\end{abstract}



\section{Introduction}

  A dissipation is normally adopted to handle discontinuities in non viscous flows (the Riemann problem). Such a dissipation can be either explicit (artificial viscosity) or implicit, being intrinsic within the adopted algorithm (e.g. Godunov-like methods). An artificial viscosity is introduced in SPH, as a shock capturing method, to prevent particle interpenetration and to smooth out spurious heating in the solution of strictly hyperbolic system of Euler equations. Such a small dissipation also produces both mass and angular momentum transport in SPH physically inviscid modelling of accretion discs \citep{a1,a6,a2,a3,a4,a12,a13,a5,a14,a10,a9,a7,a8,a11}. Efforts were accomplished in SPH to solve both the "approximate" and the "exact" Riemann problem, either via an explicit reformulation of the artificial viscosity term \citep{a15,a17,a16} or via sophisticated Godunov algorithms \citep{a18,a21,a20,a22}. A physical way to solve the Riemann problem, based on a reformulation of the EoS in the Euler equations, is here presented, where particle SPH pressure terms are recalculated without any artificial viscosity contribution. Since shock flows are non equilibrium events, we pay attention to the fact that the EoS: $p = (\gamma - 1) \rho \epsilon$ for ideal flows cannot exactly be applied to solve flow discontinuities. Applications of the reformulated EoS to 1D shock tubes \citep{a32}, to solve shocks are shown, as well as a comparison to models adopting a different dissipation is given for the coming out of spiral patterns in accretion discs in CBs.

\section{The Euler equations: SPH formulation}

  In the Lagrangian ideal non-viscous gas hydrodynamics, the relevant equations (Euler equations) are:

\begin{eqnarray}
\frac{d\rho}{dt} + \rho \nabla \cdot \underline{v} = 0 & \mbox{continuity equation} \\
\frac{d \underline{v}}{dt} = - \frac{\nabla p}{\rho} & \mbox{momentum equation} \\
\frac{d \epsilon}{dt} = - \frac{p}{\rho} \nabla \cdot \underline{v} & \mbox{energy equation} \\
p = (\gamma - 1) \rho \epsilon & \mbox{perfect gas equation} \\
\frac{d \underline{r}}{dt} = \underline{v} & \mbox{kinematic equation}
\end{eqnarray}

  The most of the adopted symbols have the usual meaning: $d/dt$ stands for the Lagrangian derivative, $\rho$ is the gas density, $\epsilon$ is the thermal energy per unit mass. The adiabatic index $\gamma$ has the meaning of a numerical parameter whose value lies in the range between $1$ and $5/3$, in principle.

 The SPH method is a free Lagrangian scheme that discretizes the fluid into moving interacting and interpolating domains called "particles". All particles move according to pressure and body forces. The method makes use of a Kernel $W$ useful to interpolate a physical quantity $A(\underline{r})$ related to a gas particle at position $\underline{r}$ according to \citep{a23,a24}:

\begin{equation}
A(\underline{r}) = \int_{D} A(\underline{r}') W(\underline{r}, \underline{r}', h) d \underline{r}' \rightarrow A_{i} = \sum_{j=1}^{N} \frac{A_{j}}{n_{j}} W(\underline{r}_{i}, \underline{r}_{j}, h) = \sum_{j=1}^{N} \frac{A_{j}}{n_{j}} W_{ij}
\end{equation}

$W(\underline{r}, \underline{r}', h)$, the interpolation Kernel, is a continuous function - or two connecting continuous functions whose derivatives are continuous even at the connecting point - defined in the spatial range $2h$, whose limit for $h \rightarrow 0$ is the Dirac delta distribution function. All physical quantities are described as extensive properties smoothly distributed in space and computed by interpolation at $\underline{r}$. In SPH terms the sum is extended to all particles included within the interpolation domain $D$, $n_{j} = \rho_{j}/m_{j}$ is the number density relative to the $jth$ particle. $W(\underline{r}_{i}, \underline{r}_{j}, h) \leq 1$ is the adopted interpolation Kernel whose value is determined by the relative distance between particles $i$ and $j$.

  In SPH formalism, equations (2) and (3) take the form, respectively:
  
\begin{equation}
\frac{d \underline{v}_{i}}{dt} = - \sum_{j=1}^{N} m_{j} 
\left( \frac{p_{i}^{\ast}}{\rho_{i}^{2}} + \frac{p_{j}^{\ast}}{\rho_{j}^{2}} \right) \nabla_{i} W_{ij}
\end{equation}

\begin{equation}
\frac{d \epsilon_{i}}{dt} = \frac{1}{2} \sum_{j=1}^{N} m_{j} \left( \frac{p_{i}^{\ast} }{\rho_{i}^{2}} + \frac{p_{j}^{\ast}}{\rho_{j}^{2}}\right)  \underline{v}_{ij} \cdot \nabla_{i} W_{ij}
\end{equation}

where $\underline{v}_{ij} = \underline{v}_{i} - \underline{v}_{j}$ and $m_{j}$ is the mass of $jth$ particle.

  For a better energy conservation, the total energy $E = (\epsilon + \frac{1}{2} v^{2})$ can also be introduced in the SPH formulation:

\begin{equation}
\frac{d}{dt} E_{i} = - \sum_{j=1}^{N} m_{j} \left( \frac{p_{i}^{\ast} \underline{v}_{i}}{\rho_{i}^{2}} + \frac{p_{j}^{\ast} \underline{v}_{j}}{\rho_{j}^{2}} \right) \cdot \nabla_{i} W_{ij}
\end{equation}

of the energy equation: $d(\epsilon + v^{2}/2)/dt = - \nabla \cdot (p \underline{v})/\rho$.

In this scheme the continuity equation takes the form:

\begin{equation}
\frac{d\rho_{i}}{dt} = \sum_{j=1}^{N} m_{j} \underline{v}_{ij} \cdot 
\nabla_{i} W_{ij}
\end{equation}

or, as we adopt, it can be written as: $\rho_{i} = \sum_{j=1}^{N} m_{j} W_{ij}$, which identifies the natural space interpolation of particle densities according to equation (6).

  The pressure terms $p^{\ast}$ include the artificial viscosity contribution given by \citet{a23,a24} and \citet{a25}, with an appropriate thermal diffusion term which reduces shock fluctuations: $p_{i}^{\ast} = p_{i} (1 + \eta_{ij})$. $\eta_{ij}$ is given by:

\begin{equation}
\eta_{ij} = \alpha \mu_{ij} + \beta \mu_{ij}^{2}, \textrm{where} \ \mu_{ij} = \left\{ \begin{array}{ll}
\frac{2 h \underline{v}_{ij} \cdot \underline{r}_{ij}}{(c_{si} + c_{sj}) (r_{ij}^{2} + \xi^{2})} & \textrm{if $\underline{v}_{ij} \cdot \underline{r}_{ij} < 0$}\\
\\
0 & \textrm{otherwise}
\end{array} \right.
\end{equation}

with $c_{si}$ being the sound speed of the $ith$ particle, $\underline{r}_{ij} = \underline{r}_{i} - \underline{r}_{j}$, $\xi^{2} \ll h^{2}$, $\alpha \approx 1$ and $\beta \approx 2$. These $\alpha$ and $\beta$ parameters of the order of the unity are usually adopted \citep{a26} to damp oscillations past high Mach number shock fronts developed by non-linear instabilities \citep{a27}. The linear $\alpha$ term is based on the viscosity of a gas. The quadratic ($\beta$, Von Neumann-Richtmyer-like) artificial viscosity term is necessary to handle strong shocks. Linear $\alpha$ and quadratic $\beta$ artificial viscosity terms are $\sim 1$. A reformulation of the artificial viscosity could be even necessary because, for "weak shocks" or low Mach numbers, the fluid becomes "too viscous" and angular momentum and vorticity could be non-physically transferred. \citet{a15,a17} developed some effective techniques to get a limitation of the effectiveness of the artificial viscosity, while \citet{a16} derived a new formulation for artificial viscosity conceptually based on the Riemann problem.

  The solution of the Riemann problem in SPH is also obtained at the interparticle contact points among particles, where a pressure and a velocity, relative to the flow discontinuity, are computed. This is also clearly shown in \citet{a21,a20,a22}, where the new pressure $p^{\ast}$ and velocity $v^{\ast}$ are reintroduced in the Euler equations to obtain the new solutions compatible with inviscid flow discontinuities. In SPH, we pay attention in particular \citep{a21,a20} to the particle pressures $p_{i}^{\ast}$ and $p_{j}^{\ast}$, in the SPH formulation of the momentum and energy equations (7) and (8 or 9), whose substitution with pressures, solutions of the Riemann problem, excludes any artificial viscosity adoption, although a dissipation is necessarily implicitly introduced \citep{a19}.

\section{How EoS matches the Riemann problem}

  The classical EoS for a perfect gas: $p = (\gamma - 1) \rho \epsilon$ is strictly applied in fluid dynamics when the gas components do not collide with each other whenever a thermal equilibrium exists. It modifies in: $p^{\star} = (\gamma - 1) \rho \epsilon +$ other in the case of gas collisions. The further term takes into account the velocity of perturbation propagation \citep{a17}. This velocity equals the ideal gas sound velocity $c_{s}$ whenever we treat static gases or in the case of rarefaction waves. Instead, it includes the "compression velocity": $\underline{v}_{ij} \cdot \underline{r}_{ij}/\mid \underline{r}_{ij} \mid$ in the case of shocks ($v_{sig, i \rightarrow j} = c_{si} - \underline{v}_{i} \cdot \underline{r}_{ij}/|\underline{r}_{ij}|$; \citet{a30}). In the first case, we write the EoS for inviscid ideal gases as: $p = \rho c_{s}^{2}/\gamma$, where $c_{s} = (\gamma p/\rho)^{1/2} = [\gamma (\gamma - 1) \epsilon]^{1/2}$. In the second case, the new formulation for the EoS is obtained squaring $v_{sig, i \rightarrow j}$, so that $c_{s}^{2} (1 - v_{shock}/c_{s})^{2}$ is an energy per unit mass in the case of shock. Hence:

\begin{equation}
p^{\star} = \left\{ \begin{array}{ll}
\frac{\rho}{\gamma} c_{s}^{2} \left(1 - \frac{v_{shock}}{c_{s}} \right)^{2} & \textrm{if $v_{shock} < 0$}\\
\\
\frac{\rho}{\gamma} c_{s}^{2} & \textrm{if $v_{shock} \ge 0$}
\end{array} \right.
\end{equation}

  In the SPH scheme, being:

\begin{equation}
p_{i}^{\star} = \frac{\rho_{i}}{\gamma} c_{si}^{2} \left(1 - \frac{v_{shock,i}}{c_{si}} \right)^{2}, \ v_{shock,i} = \left\{ \begin{array}{ll}
\frac{\underline{v}_{ij} \cdot \underline{r}_{ij}}{\mid \underline{r}_{ij} \mid} & \textrm{if $\underline{v}_{ij} \cdot \underline{r}_{ij} < 0$}\\
\\
0 & \textrm{otherwise.}
\end{array} \right.
\end{equation}

  This formulation introduces the "shock pressure term" $\rho (v_{shock}^{2} - 2 v_{shock} c_{s})/\gamma$, whose dependence on linear and quadratic power on $\underline{v}_{ij} \cdot \underline{r}_{ij}/\mid \underline{r}_{ij} \mid$ is analogue to both the linear and the quadratic components of the artificial viscosity term (eq. 11). The linear term in $\underline{v}_{ij} \cdot \underline{r}_{ij}/\mid \underline{r}_{ij} \mid$ is based on the viscosity of a gas. The quadratic term in $\underline{v}_{ij} \cdot \underline{r}_{ij}/\mid \underline{r}_{ij} \mid$ (Von Neumann - Richtmyer - like) is necessary to handle strong shocks. These contributions involve a dissipative power, whose effect correspond to an increase of the gas pressure. Therefore, we adopt the formulation (eq. 13) as $p_{i}^{\star}$ in the SPH formulation of the momentum (eq. 8) and energy equations (eqs. 9 or 10):

\begin{equation}
\frac{d \underline{v}_{i}}{dt} = - \sum_{j=1}^{N} m_{j} 
\left( \frac{p_{i}^{\star}}{\rho_{i}^{2}} + \frac{p_{j}^{\star}}{\rho_{j}^{2}} \right) \nabla_{i} W_{ij}
\end{equation}

\begin{equation}
\frac{d \epsilon_{i}}{dt} = \frac{1}{2} \sum_{j=1}^{N} m_{j} \left( \frac{p_{i}^{\star} }{\rho_{i}^{2}} + \frac{p_{j}^{\star}}{\rho_{j}^{2}}\right) \underline{v}_{ij} \cdot \nabla_{i} W_{ij},
\end{equation}

\begin{equation}
\frac{d}{dt} E_{i} = - \sum_{j=1}^{N} m_{j} \left( \frac{p_{i}^{\star} \underline{v}_{i}}{\rho_{i}^{2}} + \frac{p_{j}^{\star} \underline{v}_{j}}{\rho_{j}^{2}} \right) \cdot \nabla_{i} W_{ij}.
\end{equation}

  However, to give a generalization, we need only one general EoS and not a separation of the EoS according to the kinematic of the flow. To this purpose, we can generalize the EoS: $p^{\star} = \rho c^{2}_{s} (1 - v_{shock}/c_{s})^{2}/\gamma$ as:

\begin{equation}
p^{\star} = \frac{\rho}{\gamma} c_{s}^{2} \left(1 - C \frac{v_{R}}{c_{s}} \right)^{2},
\end{equation}

where $C \rightarrow 1$ for $v_{R} = \underline{v}_{ij} \cdot \underline{r}_{ij}/\mid \underline{r}_{ij} \mid < 0$, whilst $C \rightarrow 0$ otherwise. A simple empirical formulation can be: $C = \textrm{arccot} (R v_{R}/c_{s})/\pi$, where $R \gg 1$. $R$ is a large number describing how much the flow description corresponds to that of an ideal gas. To this purpose, $R \approx \lambda/d$, being $\lambda \propto \rho^{-1/3}$ the molecular mean free path, and $d$ the mean linear dimension of gas molecules.

\begin{figure*}
\centerline{\epsfxsize=12cm\epsfysize=8cm\epsfbox{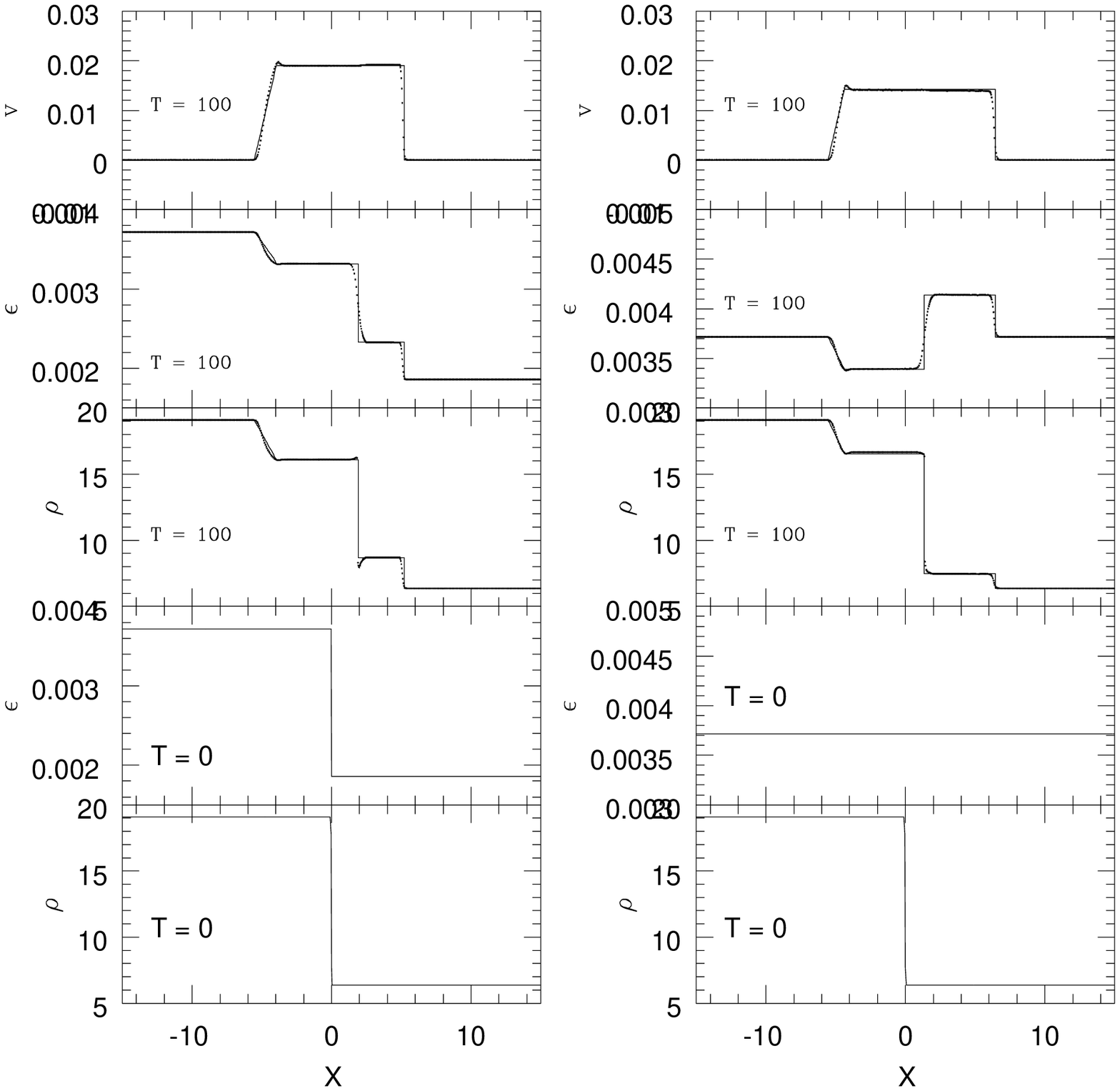}}
\caption{1D shock tube tests as far as both analytical (solid line) and our SPH-Riemann (dots) results are concerned. Density $\rho$, thermal energy $\epsilon$ and velocity $v$ are plotted at time $T = 100$. Density and thermal energy of particles initially at rest at time $T = 0$ are also reported. The initial velocity is zero throughout.}
\end{figure*}

\section{Tests: 1D Sod shock tube and 3D accretion discs in CBs}

  A comparison of analytical and our SPH - Riemann 1D shock tube test results (Sod 1978), also with the initial particle configuration (time $T = 0$), is made. Figg. 1 and 2 shows results concerning the particle density, thermal energy per unit mass and velocity at the same final computational time ($T = 100$). Throughout our SPH-Riemann simulations, the initial particle resolution length is $h = 5 \cdot 10^{-2}$. The whole computational domain is built up with $2001$ particles from $X = 0$ to $X = 100$, whose mass is different, according to the shock initial position. At time $T = 0$ all particles are motionless. $\gamma = 5/3$, while the ratios $\rho_{1}/\rho_{2} = 3$ and $\epsilon_{1}/\epsilon_{2} = 2$ (Fig. 1 - left side), and  $\rho_{1}/\rho_{2} = 3$ and $\epsilon_{1}/\epsilon_{2} = 1$ (Fig. 1 - right side), between the two sides left-right. The first and the last $5$ particles of the 1D computational domain, keep fixed positions and do not move. Discrepancies involve only $4$ particles at most. This means that the physical dissipation introduced in the EoS (eq. 13) is effective.

\begin{figure*}
\centerline{\epsfxsize=12cm\epsfbox{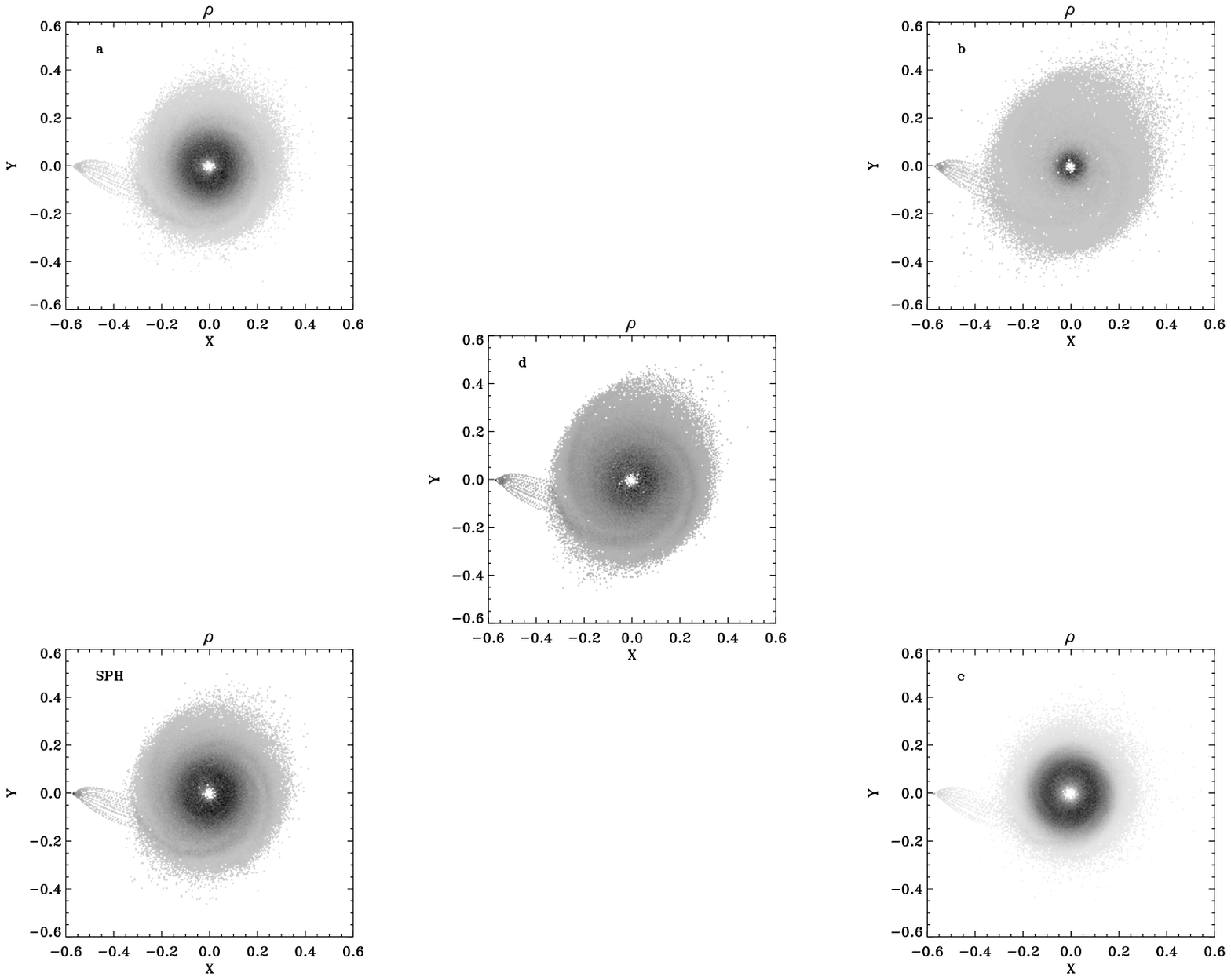}}
\caption{XY plots of $64$ greytones density $\rho$ isocontours of 3D disc modelling in CBs. SPH refers to normal SPH results. Disc model "a" refers to SPH modelling where \citet{a15} attenuation is introduced. Model "b" refers to SPH modelling where \citet{a17} treatment of the Riemann solver is introduced. Model "c" refers to SPH modelling where \citet{a16} $\alpha$ decay is introduced. Model "d" refers to SPH modelling where the EoS reformulation is introduced.}
\end{figure*}

  Results on 3D SPH simulations of accretion disc structure and dynamics in a CB are shown as far as the coming out of spiral patterns and shocks in the radial flow are concerned. The resolution length of SPH particles is $h = 0.005$, being $1$ the non-dimensional separation $d_{12} = 10^{6} \ Km$ of the two stars. The injection of particles from the inner Lagrangian point L1 is supersonic: $v_{inj} = 130 \ Km \ s^{-1}$, whilst the temperature of gas coming from the secondary star is $T = 10^{4} \ K$ and $\gamma = 1.01$. The compact primary is a $1 M_{\odot}$ star, while the donor companion is a $0.5 M_{\odot}$ star. The comparison with other different SPH techniques (see Fig. 2 caption), where artificial dissipation is explicitly introduced, shows that the coming out of spirals, as well as other disc geometric details, are much better evidenced in those simulations where the EoS and related dissipation are treated in their full physical sense.

  3D accretion disc simulations clearly show different results. The adoption of a supersonic injection kinematics from L1 is justified in terms both of conservation of the flux momentum and in terms of the Jacobi constant and in terms of the Bernoulli's theorem \citep{a44}. This kinematics concern active phases of accretion discs where both radial disc increases and spiral structures develop. Causes generating spiral patterns and disc geometric details are deeply discussed in \citet{a34,a36,a37,a35,a38,a12,a13} as a consequence of external perturbations as tidal torques as well as the collisional impact of the injected flow onto the disc's outer edge. Hence, the 3D simulated structure better corresponding to such theoretical and observational constraints is that produced according to the formalism here proposed (model d). The typical 3D SPH disc structure (model a) does not fully correspond to these characteristics, while other 3D disc models lack of several features.




\end{document}